\documentstyle[multicol,prl,aps,epsfig]{revtex}

\begin{document}

\draft
\title{Microwave Electrodynamics of Electron-Doped Cuprate Superconductors}

\author{J. David Kokales, Patrick Fournier, Lucia V. Mercaldo, Vladimir V. Talanov, Richard L. Greene, and Steven M. Anlage}
\address{Center for Superconductivity Research, Department of Physics, University of Maryland, College Park, MD 20742-4111}
\maketitle

\begin{abstract}
We report microwave cavity perturbation measurements of the temperature dependence of the penetration depth, $\lambda$(T), and conductivity, $\sigma$(T) of Pr$_{2-x}Ce_{x}CuO_{4-\delta}$ (PCCO) crystals, as well as parallel-plate resonator measurements of $\lambda$(T) in PCCO thin films.  Penetration depth measurements are also presented for a Nd$_{2-x}Ce_{x}CuO_{4-\delta}$ (NCCO) crystal.  We find that $\Delta\lambda$(T) has a power-law behavior for $T<T_c/3$, and conclude that the electron-doped cuprate superconductors have nodes in the superconducting gap.  Furthermore, using the surface impedance, we have derived the real part of the conductivity, $\sigma_1$(T), below T$_c$ and found a behavior similar to that observed in hole-doped cuprates.
\end{abstract}

\begin{multicols}{2}
\columnseprule 0pt
\narrowtext
Existing experimental data on the electron-doped cuprate superconductors \cite{Tokura} have been interpreted as being consistent with an s-wave pairing state symmetry \cite{Fournier}.  In particular, magnetic screening length measurements versus temperature have been interpreted as showing an activated behavior consistent with s-wave symmetry \cite{Wu,Andreone,Schneider,Anlage94,Alff}, mainly with a large activation gap, 2$\Delta(0)/k_BT_c >$ 4 \cite{Wu,Andreone,Schneider,Anlage94}.  This is in contrast to the predominantly d-wave behavior widely observed in the hole-doped cuprates \cite{Hardy}, most conclusively demonstrated in phase-sensitive experiments \cite{VanHar}.  Raman scattering supports a d-wave symmetry in the hole-doped cuprate superconductors \cite{Devereaux} and suggests s-wave in electron-doped cuprates \cite{Stadlober}.  Tunneling spectroscopy measurements have also been interpreted in the same manner \cite{Alff,Huang,Alff98}.  A zero-bias conductance peak is expected for tunneling into the gap nodal directions in d-wave superconductors due to the change in sign of the order parameter on the Fermi surface, allowing for the formation of Andreev bound states at the Fermi energy \cite{Hu}.  Such a peak has been seen in such superconductors as $YBa_2Cu_3O_{7-\delta}$ (YBCO) \cite{Alff98}, which are known to have a dominant d-wave order parameter.  Until recently \cite{Hayashi}, no such peak was observed in the electron-doped cuprates \cite{Alff,Alff98}, consistent with an s-wave symmetry for these materials. 

Since there is no distinction between electron and hole doping of the Hubbard model around half-filling \cite{Nagaoka,Emery}, one does not expect electron- and hole-doped cuprates to have different pairing mechanisms and symmetries.  In this letter we present evidence for nodes in the excitation spectrum, and by implication that the pairing state symmetry in the electron-doped cuprates is predominantly d-wave.  This evidence comes from temperature dependent penetration depth measurements which, although not able to singularly determine the pairing state symmetry, provide strong indications that there are an abundance of low energy excitations in these materials.

The strongest data supporting s-wave symmetry has come from temperature dependent measurements of the magnetic screening length in NCCO \cite{Wu,Andreone,Schneider,Anlage94}.  However, there were significant limitations of these experiments.  First, the paramagnetism present in the rare-earth ions of these materials may have influenced the behavior of the deduced penetration depth as a function of temperature, and thus the determination of the pairing state symmetry, as described by Cooper, {\it et al.} \cite{Cooper} and demonstrated by Alff, {\it et al.} \cite{Alff}.  Secondly, the materials may not have been studied to sufficiently low temperature to sort out the influence of paramagnetism in the screening length data \cite{Wu,Andreone,Schneider,Anlage94}.  Finally, the influence of paramagnetism on the screening length measurement in those experiments was strongest for the chosen orientation of the samples and the rf field (H$_{rf} \perp$ c-axis).  In addition, some of the samples were measured under conditions in which both a-b plane and c-axis currents were stimulated.  This results in measurements of a weighted sum of $\lambda_{ab}$ and $\lambda_c$.

In this work we have addressed all of these concerns.  First, we studied Pr$_{1.85}Ce_{0.15}CuO_{4-\delta}$ (PCCO), in which the Pr ion has a much smaller, and less temperature dependent, paramagnetism than that present in the previously studied NCCO \cite{Dali}.  Secondly, the orientation ($H_{rf} \parallel$ c) of the rf field in our microwave cavity is such that only the Cu-O plane screening currents are stimulated, making it possible to extract the intrinsic in-plane penetration depth without having to account for c-axis effects.  This orientation also significantly reduces the effects of the RF paramagnetism on the screening length measurements \cite{Dali}.  Finally, we did our measurements down to 1.2 K, a significantly lower temperature than all previous experiments.  Other significant improvements to our measurement technique are discussed elsewhere \cite{M2S}.  These improvements enable us to make measurements of the temperature dependence of the penetration depth more accurately and with fewer extrinsic influences than before.

Measurements were done on both single crystals and thin films of PCCO, and an NCCO single crystal.  The directional solidification technique used to grow the single crystal samples, and their normal state physical properties, have been discussed elsewhere \cite{Peng,Brinkman}.  Typically, the PCCO samples exhibited a mid-point transition temperature of 19 K with a transition width of 1.5 K as determined by resistivity measurements, and residual normal state resistivity values of about 60 $\mu\Omega$-cm.  A typical crystal size was 1mm x 1mm x 30$\mu$m, sufficiently thin to achieve homogeneous Ce-doping in the c-direction of the crystal \cite{Skelton}.  More than 20 crystals were measured during the course of this work.  The PCCO thin films, 400 nm thick, were grown by pulsed laser deposition on LaAlO$_3$ substrates \cite{Maiser}.

Data will be presented from two different experimental techniques.  The crystals were measured by the cavity perturbation method in  a superconducting cylindrical niobium cavity operating in the TE$_{011}$ mode at 9.6 GHz.  The sample is supported on a sapphire hot finger, and its temperature can be elevated while the cavity remains at a temperature of 1.2 K \cite{Sridhar}.  The experiment involves measuring shifts in the resonant frequency, $\Delta\omega$(T), and the quality factor, Q(T), of the system as the temperature of the sample is varied \cite{Anlage94}.  This data can then be converted, using the geometry factor, $\Gamma$, into the surface resistance, R$_S$, and change in the penetration depth, $\Delta\lambda$(T), as a function of the temperature using the following relationships: $\Delta\lambda(T) = (2\Gamma/\mu_o\omega^2)(\Delta\omega_{exp}(T)-\Delta\omega_{bck}(T))$ and R$_s$(T) = $\Gamma(1/Q_{exp}(T)-1/Q_{bck}(T))$, where $\Delta\omega_{exp}$(T) and Q$_{exp}$(T) are the frequency shift and quality factor with the sample present, and $\Delta\omega_{bck}$(T) and Q$_{bck}$(T) are the background frequency shift and quality factor.  In order to reduce noise in the data, averaging of the transmission response of the cavity was performed, and two measurements were made at each temperature in a sweep.  Approximately five sweeps would be taken in immediate succession to perform further averaging and to determine a standard deviation for each data point.  This standard deviation is represented by the error bars in the figures, although they are often smaller than the data point symbol.  The PCCO film was measured using a parallel plate resonator to determine the change in the (finite thickness corrected) effective penetration depth and surface resistance as a function of temperature \cite{Taber,Talanov}.

In Fig. 1 the measured changes in the low temperature penetration depth are presented for two PCCO crystals, an NCCO crystal, and a pair of PCCO films.  Several things are of note regarding this data.  First, the low temperature upturn observed in the NCCO crystal is indicative of the paramagnetic influence on the measured magnetic screening length \cite{Alff,Cooper,Prozorov}.  Since previous experiments \cite{Andreone,Schneider,Anlage94} only went down to about 3.0 K (about 0.12 T$_c$), too high of a temperature to clearly observe the upturn, it is apparent that these determinations of the magnetic screening length temperature dependence in NCCO could have been corrupted.  Secondly, this paramagnetic upturn is absent in both the PCCO crystals (H$_{rf}\parallel$c) and the PCCO film (H$_{rf}\perp$c) data.  This demonstrates that paramagnetism does not have a significant influence on our measurements of the screening length in PCCO, as expected.  

\begin{figure}
\begin{center}
\leavevmode
\epsfig{file=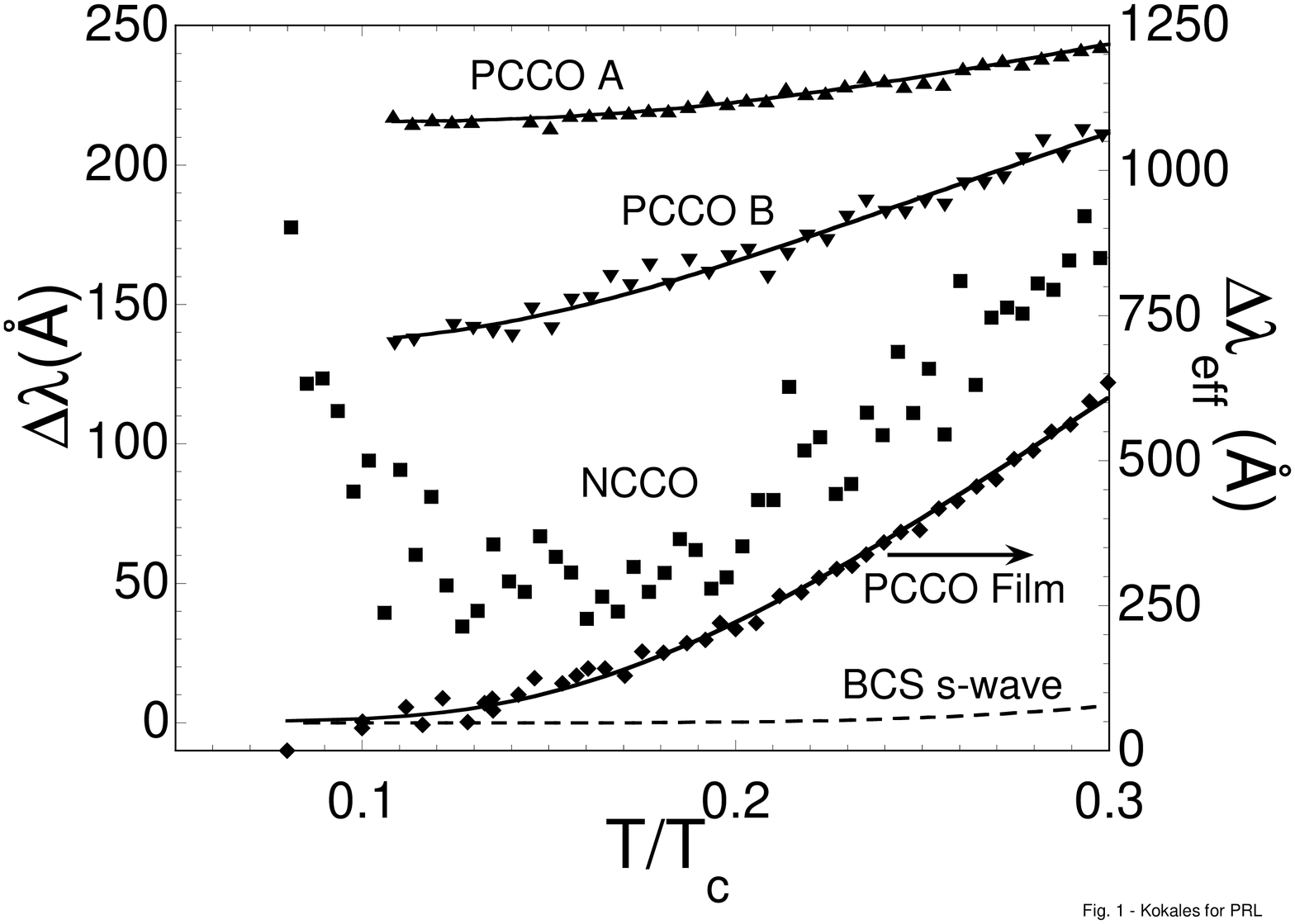,width=8.5cm,clip=}
\end{center}
\caption{Change in penetration depth, $\Delta\lambda(T/T_c)$, for: PCCO crystals A (upright triangle) and B (inverted triangle), NCCO crystal (square), $\Delta\lambda_{eff}(T/T_c)$ for PCCO films (diamond), and the BCS s-wave model for $\lambda$(0) = 1500 \AA\ and $2\Delta(0)/k_BT_c$ = 4.0 (dashed line).  Note: $\Delta\lambda$ curves have been offset for clarity, and PCCO film data is finite thickness enhanced $\Delta\lambda_{eff}$.  The solid lines show BCS s-wave fits of $\Delta\lambda(T/T_c)$ for $T/T_c<0.3$.  Parameter values for fits are given in Table I.}
\label{schematic}
\end{figure}

It is possible to examine the low energy excitation spectrum by carefully analyzing $\Delta\lambda$(T) at low temperatures, T$<T_c$/3, without making any assumptions about the value of $\lambda$(0).  For clean line nodes in the gap, a linear behavior for $\Delta\lambda$(T) is expected, whereas a quadratic behavior is expected if these nodes are filled by impurity states \cite{Annett,Hirschfeld}.  For convenience, we shall refer to these cases as $\lq\lq$clean-" and $\lq\lq$dirty-"nodes respectively.  In the case of BCS s-wave symmetry, an activated form of $\Delta\lambda$(T), given by $\Delta\lambda(T) = \lambda(0)(\pi\Delta(0)/2k_BT)^{1/2}e^{-\Delta(0)/k_BT}$ for $T << T_c/2$, is expected \cite{Anlage94}.  Considering first the possibility of s-wave behavior in PCCO, the data on all samples in Fig. 1 can be fit to an s-wave activated behavior, but only with unphysically small $\Delta$(0) and $\lambda$(0) values - see Table I.  After measuring many samples, most not shown here, it was found that those samples which exhibit the lowest surface resistances typically have small fit gap values, 2$\Delta(0) < 1.9 k_BT_c$.    In contrast, Alff {\it et al.} \cite{Alff} found 2$\Delta$(0) = 2.9 k$_B$T$_c$ from screening length measurements in PCCO films.  Clearly, we must conclude that the screening length temperature dependence of PCCO is not consistent with the clean s-wave behavior with 2$\Delta(0)/k_BT_c \ge$ 4, found earlier in NCCO \cite{Wu,Schneider,Anlage94}.

\pagebreak
\end{multicols}
\widetext
\begin{table}
\caption{Fitting parameters for $\Delta\lambda$(T) (T$ \le T_c$/3) for BCS s-wave, clean d-wave, and dirty d-wave models.  $T_c$ and $\Delta$$T_c$ values, estimated from surface resistance data for the crystals, and ac susceptibility for the films, are also shown.  The parameter error estimates are those required to double the value of $\chi^2$.  Note that the s-wave fits have 4 parameters, while the d-wave fits have 3 parameters.}
\begin{tabular}{|rrrrrrrrrr|}
\hline
 \vline\vline& & \vline\vline& BCS & s-wave & \vline\vline& clean & d-wave \vline\vline& dirty & d-wave \\ 
\hline
Sample \vline\vline& $T_c$ \vline& $\Delta$$T_c$ \vline\vline& $\lambda(0)$(\AA) \vline & 2$\Delta(0)/k_BT_c$ \vline& $\chi^2$ \vline\vline& c$_1$(\AA/K) \vline& $\chi^2$ \vline\vline& c$_2$(\AA/K$^2$) \vline& $\chi^2$ \\ 
\hline
\hline
PCCO crystal A \vline\vline& 19.0 \vline& 1.5 \vline\vline& 295 \vline& 1.89 $\pm$ 0.06 \vline& 2.53 \vline\vline& 8.0 $\pm$ 0.7 \vline& 7.22 \vline\vline& 1.0 $\pm$ 0.1 \vline& 3.54 \\ 
\hline
PCCO crystal B \vline\vline& 19.0 \vline& 1.7 \vline\vline& 390 \vline& 1.37 $\pm$ 0.05 \vline& 16 \vline\vline& 21 $\pm$ 1 \vline& 19.3 \vline\vline& 2.7 $\pm$ 0.2 \vline& 14.5 \\ 
\hline
PCCO Film \vline\vline& 20.0 \vline& 0.5 \vline\vline& 2660 \vline& 1.62 $\pm$ 0.04 \vline& 230 \vline\vline& 115 $\pm$ 8 \vline& 1300 \vline\vline& 15 $\pm$ 0.7 \vline& 225 \\ 
\hline
NCCO crystal \vline\vline& 24.0 \vline& 1.5 \vline\vline& - \vline& - \vline& - \vline\vline& - \vline& - \vline& - \vline& - \\
\hline
\end{tabular}
\end{table}
\begin{multicols}{2}
\columnseprule 0pt
\narrowtext

\begin{figure}
\begin{center}
\leavevmode
\epsfig{file=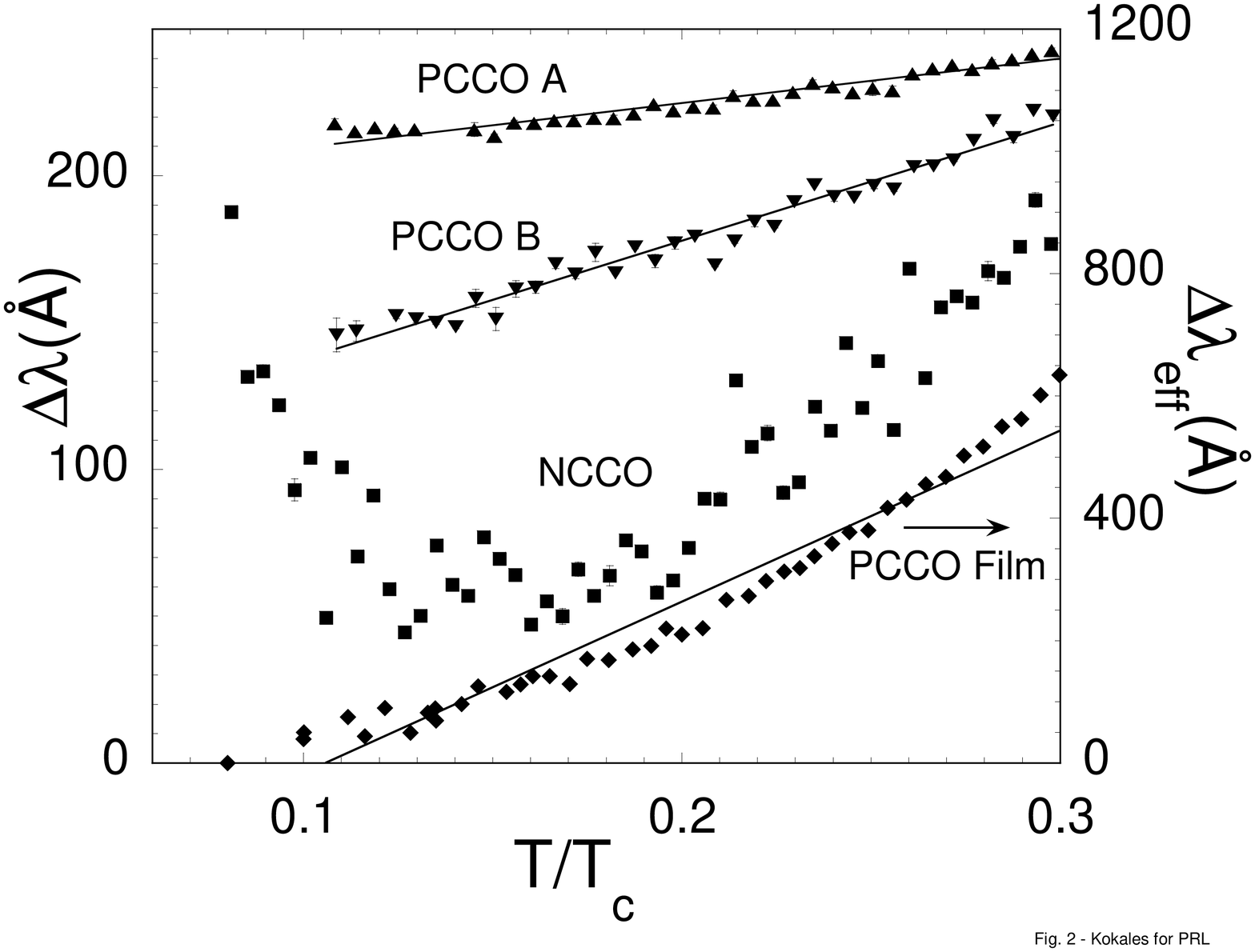,width=8.5cm,clip=}
\end{center}
\caption{Change in penetration depth vs T/T$_c$ with clean node fits (solid lines) to $\Delta\lambda$(T) = $c_1T + offset$.  See Fig. 1 for legend.  Parameter values for fits are given in Table I.}
\label{schematic}
\end{figure}

In Figs. 2 and 3 fits to linear and quadratic temperature dependencies, expected for clean and dirty nodes respectively, are presented.  These fits describe the data quite well and generate reasonable values for the temperature dependence prefactors. One expects clean d-wave nodes to exhibit $\Delta\lambda$(T) = c$_1$T+b \cite{Hirschfeld} with c$_1$ = k$_B\lambda(0)ln(2)/\Delta(0)\cong$ 28 \AA/K, using $\lambda$(0) = 1500$\ \AA$ \cite{Nugroho} and 2$\Delta$(0)/k$_BT_c$ = 3.9 \cite{Huang}.  For a dirty d-wave node, we expect $\Delta\lambda$(T) = c$_2T^2$+d with c$_2 \cong \lambda(0)/(\Gamma^{1/2}\Delta^{3/2}(0)) \sim$ 10 \AA/K$^2$, where $\Gamma$ is the (unitary limit) scattering rate, which is proportional to the impurity concentration of the sample \cite{Hirschfeld}.  One expects that a high value for $\Gamma$ leads to a lower value for $T_c$ due to impurity scattering.  This was verified in our quadratic fits, which revealed a general trend towards higher scattering rates as T$_c$ decreased.  The order-of-magnitude value for c$_2$ is arrived at by scaling the observed value of c$_2$ = 0.7 \AA/K$^2$ from Bi$_2Sr_2CaCu_2O_{8+\delta}$ (Bi2212) \cite{Ma} by the ratio of the gap values between Bi2212 and PCCO.

For fits to a clean d-wave node behavior, we find the PCCO crystals have an average value for c$_1$ of 24 \AA/K, which is close to the theoretical estimated value.  However some of the samples, such as the PCCO films, exhibit a higher power-law temperature dependence (Fig. 2).  When fit to the T$^2$ functional form, the PCCO films gave c$_2$ = 15 \AA/K$^2$, the expected order of magnitude for c$_2$, suggesting a dirty d-wave node behavior.  Overall, we find that much of our data is consistent with a dirty d-wave form, showing c$_2$ values ranging from 1.0 to 15 \AA/K$^2$, although the linear fits are not significantly worse in some cases.  This can be verified by comparison of the $\chi^2$ values for the three fits given in Table I.  It should be noted that our results for $\Delta\lambda$(T) are in good agreement with those taken by an LC resonator method on similar PCCO crystals by Prozorov {\it et al} \cite{Prozorov}.  Both groups see an upturn in $\Delta\lambda$(T) at low temperatures in NCCO, as well as power-law behavior for $\Delta\lambda$(T) in PCCO, with quantitative agreement for $\Delta\lambda$ (up to a geometry factor difference) on a common crystal.

\begin{figure}
\begin{center}
\leavevmode
\epsfig{file=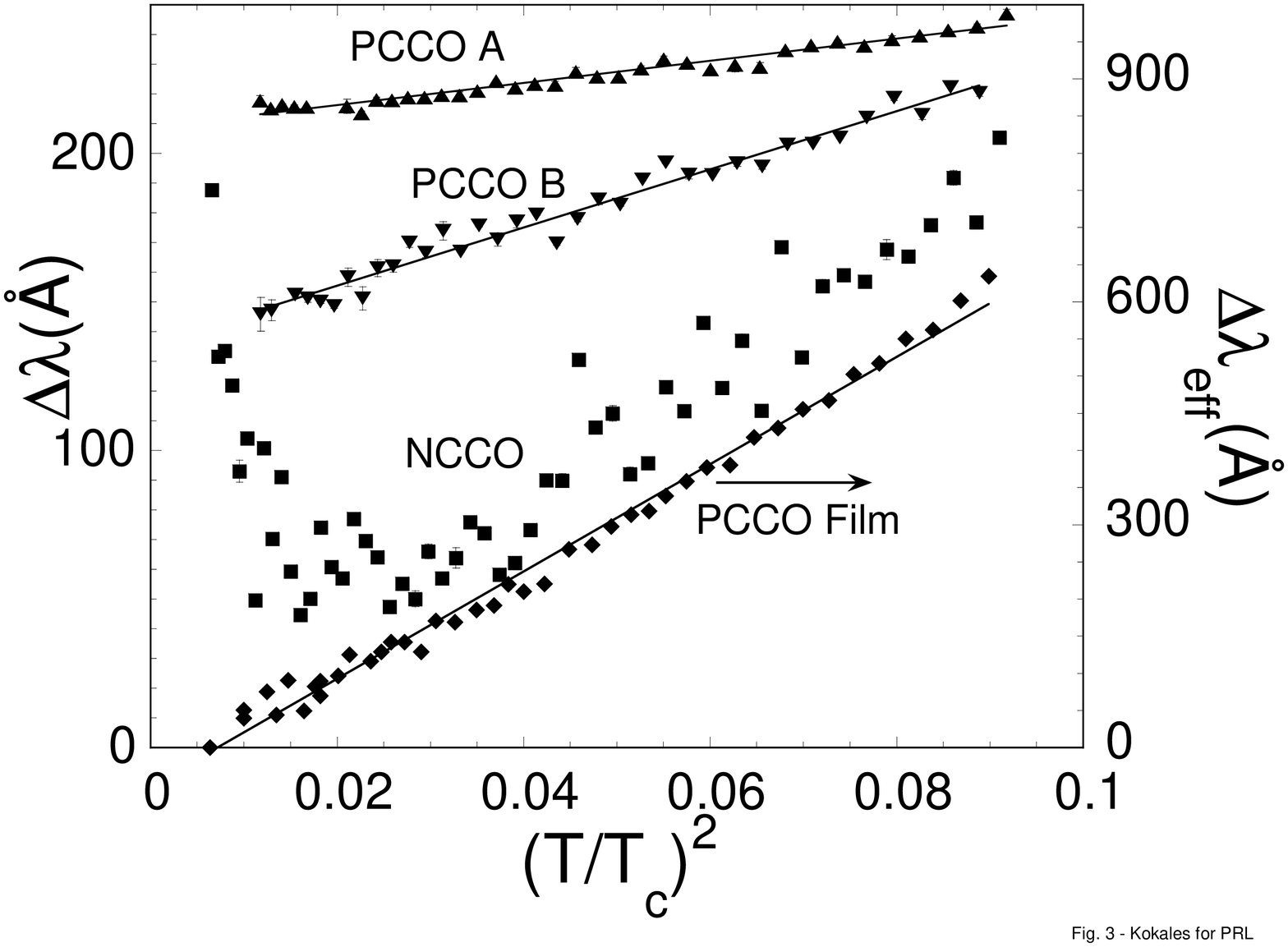,width=8.5cm,clip=}
\end{center}
\caption{Change in penetration depth vs (T/T$_c$)$^2$ with dirty node fits (solid lines) to $\Delta\lambda$(T) = $c_2T^2 + offset$.  See figure 1 for legend.  Parameter values for fits are given in Table I.}
\label{schematic}
\end{figure}

Although our data generally indicates a power-law behavior for the screening length temperature dependence, making d-wave symmetry a prime candidate, it is possible that a gapless s-wave symmetry could also be responsible for a T$^2$ behavior \cite{Abrahams}.  However, the fact that $\Delta\lambda \sim T$ for some of the PCCO crystals, with reasonable prefactors, indicates that d-wave symmetry is the most likely explanation.  After this paper was submitted, we became aware of phase-sensitive measurements on NCCO and PCCO thin films which concluded that these materials have d-wave symmetry \cite{Tsuei}.

Because distinct multiple transitions are not evident in our data (see R$_S$(T) in fig. 4), there is not a clear mixed pairing state symmetry in these materials \cite{Annett}.  However, the observation of c-axis Josephson supercurrents between Pb and NCCO \cite{Woods} suggests that a sub-dominant s-wave order parameter may exist in the electron-doped cuprates.

We can also determine the real part of the complex conductivity, $\sigma_1$(T), of these crystals from our measurements.  Fig. 4 presents $\sigma_1$(T) for two of the PCCO crystals, obtained using the local limit expression Z$_s = (i\omega\mu_o/\sigma)^{1/2}$ and setting X$_S$(T) equal to R$_S$(T) at T=30 K.  This data presents some similarities to data on YBCO \cite{Bonn}\cite{Anlage96}, and is distinctly different from the BCS coherence peak (calculated with T$_C$ = 20 K, $\xi_{BCS}$ = 80 \AA, l$_{MFP}$ = 200 \AA, and $\Delta$(0)/k$_B$T$_C$ = 1.76 \cite{Halbritter}).  Our results for the electron-doped cuprates are similar to those on Bi2212 \cite{Lee} and exhibit behavior similar to YBCO at temperatures above 0.4T$_c$.  It should be noted that the PCCO data has not had any residual resistance subtracted.  An extrinsic residual resistance will alter the form of $\sigma_1$(T), particularly at low temperatures (see the NCCO data in Fig. 4).  Furthermore, the magnitude of $\sigma_1$(T) is sensitive to the choice of $\lambda$(0).  However, we found the shape of the $\sigma_1$(T)/$\sigma_n$ curves to be insensitive to the chosen $\lambda$(0) value in the range of 1150 \AA\ to 4720 \AA.  Thus, irregardless of the choice of $\lambda$(0) and residual resistance, the conductivity temperature dependence is qualitatively similar to YBCO \cite{Anlage96} or Bi2212 \cite{Lee}, suggesting that quasiparticle dynamics in the electron-doped cuprates are similar to those of the hole-doped cuprates.  These similarities continue into the normal state (H $>$ H$_{c2}$) where the conductivity of PCCO thin films is found to be qualitatively similar to that of $La_{2-x}Sr_xCuO_{4-\delta}$ \cite{Fournier98}.

\begin{figure}
\begin{center}
\leavevmode
\epsfig{file=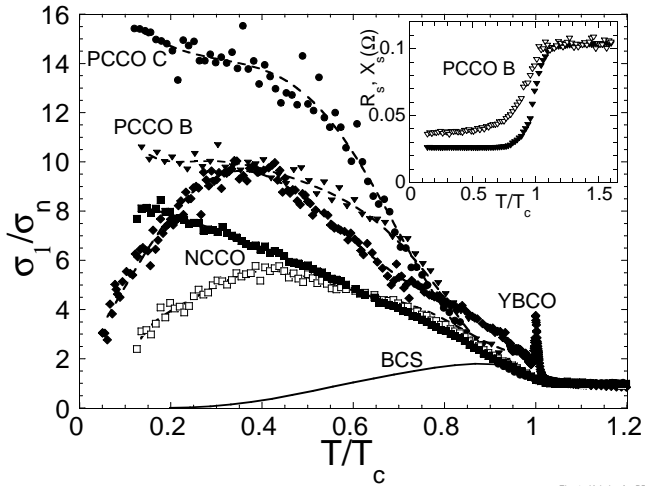,width=8.5cm,clip=}
\end{center}
\caption{Real part of the conductivity, $\sigma_1(T)$, normalized to its normal state value $\sigma_n$ at 9.6 GHz for: PCCO crystals B (inverted triangle) and C (circle), an NCCO crystal with (open square) and without (closed square) a residual resistance (9.6 m$\Omega$) subtracted, and a YBCO crystal (diamond) \protect\cite{Anlage96}.  Dashed lines are guides to the eye.  Solid line is the BCS coherence peak.  Inset:  R$_S$(T) (closed) and X$_S$(T) (open) for PCCO B at 9.6 GHz over the full temperature range.}
\label{schematic}
\end{figure}

In summary, we have re-examined the electrodynamic properties of the electron-doped cuprate superconductors.  We have chosen a system with less paramagnetism and made significant improvements in the experiment compared to prior work.  Given the inability to understand the $\Delta\lambda$(T) data in an s-wave picture, one must conclude that the screening length data on the electron-doped cuprate superconductors are not consistent with a large isotropic gap in the quasiparticle excitation spectrum.  We have found the temperature dependence of the penetration depth for the PCCO crystals and films to behave in a linear or quadratic manner for T/T$_c < 0.3$.  This is indicative of a node in the superconducting gap and is consistent with a d-wave pairing state symmetry.  Finally, the real part of the complex conductivity is reminiscent of that observed in YBCO and Bi2212.  This suggests that the quasiparticle dynamics and the superfluid response are qualitatively similar in electron-doped and hole-doped cuprates.

This work was supported by NSF DMR-9732736, NSF DMR-9624021, and the Maryland Center for Superconductivity Research.  We acknowledge discussions with D. H. Wu and R. Prozorov.

\end{multicols}

\end{document}